\documentstyle[prl,aps]{revtex}

\begin{document}

\twocolumn[\hsize\textwidth\columnwidth\hsize\csname
@twocolumnfalse\endcsname
\preprint{ }

\title{Minimal Supersymmetric Left-Right Model}

\author{
Charanjit S. Aulakh$^{(1)}$, Alejandra Melfo$^{(2)}$ and
Goran Senjanovi\'c$^{(3)}$}

\address{$^{(1)}$ {\it Dept. of Physics, Panjab University,
Chandigarh, India   }}
\address{$^{(2)}$ {\it International School for Advanced Studies, Trieste,
 Italy {\rm and} CAT, Universidad de
 Los Andes, M\'erida, Venezuela }}
\address{$^{(3)}${\it International Center for Theoretical Physics,
 Trieste, Italy }}

\maketitle

\begin{abstract}
We construct  the minimal left-right symmetric model by utilizing only
the fields  dictated by supersymmetry and automatic R-parity conservation.
Allowing for  non-renormalizable operators in the superpotential,  we show
that parity can  be spontaneously broken while preserving electromagnetic
gauge invariance.  The scale of parity breakdown is
predicted in the  intermediate region: $M_R > 10^{10}- 10^{11} GeV$, and
 R-parity remains exact.  The theory contains a number of
charged and doubly charged Higgs  scalars with a low mass of order
$M_R^2/M_{Planck}$, accessible to experiment.

\end{abstract} 
  ]
\vspace{0.3cm}

{\bf A.  Introduction.} \hspace{0.5cm} Certainly the most popular extension
 of the minimal Standard Model is its supersymmetric counterpart. Other
 very popular ones are Left-Right symmetric theories \cite{ps74}, 
which attribute the 
observed parity asymmetry in the weak interactions to the spontaneous 
breakdown of Left-Right symmetry, i.e. generalized parity transformations.
 Furthermore, Left-Right symmetry plays an important role in attempting to
 understand the smallness of CP violation \cite{bt78}, and in this sense
 provides an
 alternative to Peccei-Quinn symmetry.

Recently it has been pointed out that a particularly simple solution to the
 strong CP problem results from the merging of these two proposals
\cite{mr96}. Another,
 maybe more important {\it raison d'etre} for supersymmetric Left-Right
 models is the fact that they lead naturally to R-parity conservation.
Namely, Left-Right models contain a $B-L$ gauge symmetry, which allows
for this possibility \cite{m86}. All that is  needed is that one uses
a version  of the theory that incorporates a see-saw mechanism \cite{seesaw}
at the renormalizable level. More precisely, R-parity (which keeps
particles invariant, and changes the sign of  sparticles) can be
written as
\begin{equation}
R = (-1)^{3 (B - L) + 2 S}
\label{rparity}
\end{equation}
where $S$ is the spin of the particle. It can be shown that in these
kind of theories, invariance under $B-L$ implies R-parity conservation
\cite{m86}.

However, the construction of specific models turned out to be unexpectedly
 non-trivial. Namely, in the minimal version of the theory, at the
 renormalizable level, symmetry breaking is not possible
\cite{km94}. This may be
 cured by adding more fields to the theory \cite{abs} and/or assuming that the
 scale ($M_R$) of Left-Right symmetry breaking is not greater than 
the scale of supersymmetry breaking \cite{km94}. 
 We should  mention that phenomenological aspects of the supersymmetric   
left-right  theories were also studied in \cite{lrphen}, without worrying
about the  problem of symmetry breaking.  We address our attention
precisely to this, the central issue of the theory.

Although the experiments still allow for  a light $M_R$, we take seriously 
the possibility of a large $M_R$ scale, as hinted by both the 
phenomenological success of the Standard Model and neutrino physics.
In such case, the only hope for a realistic theory lies in  considering
 higher-dimensional operators. This is the scope of this letter.

Using non-renormalizable operators, we construct the minimal supersymmetric
 Left-Right model, and show that it naturally can account for 
spontaneous breakdown of parity. Furthermore, the electromagnetic charge
 and color-preserving minimum also automatically leads to an exact R-parity, 
even after integrating the large scale $M_R$ out. As is well known, 
preserving R-parity implies the stability of the lightest supersymmetric
 particle, which has well defined phenomenological implications and provides
a dark matter candidate.

It is interesting to compare the theory with the minimal renormalizable 
supersymmetric Left-Right model \cite{abs}. First, in this case  naturalness
 demands that $M_R$ be bigger  than about $10^{10} GeV$. Furthermore, 
there is an important difference in the implementation of the see-saw 
mechanism, since here, as much as in the minimal 
non-supersymmetric models,
 the  mechanism does not stay in its minimal form. We discuss this
 in detail below.

A main feature of these models is the presence of a small
 scale $m\sim M_R^2/M_{Planck}$. A number of Higgs particles, specially
 charged and doubly-charged ones have their mass proportional to $m$. This
 provides the central phenomenological implication of the theory, since 
for $M_R$ in the intermediate regime $10^{10} - 10^{12} GeV$, relevant for
 neutrino physics, these particles became accessible to new accelerators.
 This is perhaps the most appealing aspect of the theory. We proceed 
now with the construction of the model.

\vspace{0.3cm}

{\bf B. The Minimal Renormalizable Model.} \hspace{0.5cm}
The minimal Left-Right extension to the
 Standard Model \cite{ps74}
 is based on the gauge group $SU(2)_L\times SU(2)_R\times U(1)_{B-L}$. 
Its supersymmetric version contains three generations of quark and 
lepton chiral superfields
transforming as:
\begin{eqnarray}
Q=(3,2,1,1/3)&\;\;\;\;\;  & Q_c=(3^*,1,2,-1/3) \nonumber \\ 
L=(1,2,1,-1)& \;\;\;\;\; & L_c=(1,1,2,1) 
\label{matter}
\end{eqnarray}
where the numbers in the brackets denote the quantum numbers under
$SU(3)_c$, $SU(2)_L$, $ SU(2)_R$ and $ U(1)_{B-L}$ respectively.

The Higgs sector consists of two left-handed and two right-handed triplets
\begin{eqnarray}
  \Delta=(1,3,1,2)  , \quad 
  \bar{\Delta} =(1,3,1,-2) \nonumber \\
\Delta_c=(1,1,3,-2), \quad \bar{\Delta_c} =(1,1,3,2)
\label{higgs1}
\end{eqnarray}
in charge of breaking $SU(2)_R$ symmetry at a large scale $M_R$; the
 choice of the adjoint representation is the minimal necessary to achieve
 a see-saw mechanism for the neutrino mass, and the number of fields is
 doubled with respect to the non supersymmetric version to ensure anomaly
 cancellations. Of course, one could achieve a see-saw mechanism
through non-renormalizable operators even if one uses doublets instead
of triplets. However, in this case just as in the MSSM one looses
R-parity. 

 Likewise, to break the remaining Standard Model symmetry
 two bidoublets are necessary

\begin{equation}
 \Phi_i=(1,2,2,0) \quad (i = 1, 2)
\label{higgs2}
\end{equation}
with $i= 1, 2$,  in order to achieve a
 nonvanishing CKM quark mixing matrix.

The gauge symmetry is augmented by a discrete parity or Left-Right (L-R) 
symmetry under which the fields transform as
\begin{eqnarray}
Q              \leftrightarrow     Q_c^* ,\quad 
L              \leftrightarrow     L_c^* ,\quad
\Phi_i         \leftrightarrow     \Phi_i^\dagger,  \quad 
\Delta         \leftrightarrow  \Delta_c^* ,\quad
\bar{\Delta}   \leftrightarrow   \bar\Delta_c^* \nonumber .
\end{eqnarray}

With this Higgs content, the most general {\it renormalizable}
superpotential is given by

\begin{eqnarray}
 W_0 &=&m  {\rm  Tr}\, \Delta \bar{\Delta}
       + m^*  {\rm Tr}\,\Delta_c \bar{\Delta_c} \nonumber 
+ \mu_{ij} {\rm Tr}\,  \tau_2 \Phi^T_i \tau_2 \Phi_j \\
&  &+i {\bf f} L^T \tau_2 \Delta L+i {\bf f}^* L_c^T\tau_2 \Delta_c L_c 
\nonumber \\
& &+ {\bf h}_l^{(i)} L^T \tau_2 \Phi_i \tau_2 L_c
+ {\bf h}_q^{(i)} Q^T \tau_2 \Phi_i \tau_2 Q_c 
\label{superpot}
\end{eqnarray}

where ${\bf h}^{(i)}_{q,l}  =  {{\bf h}^{(i)}_{q,l}}^\dagger $,
$\mu_{ij}  =  \mu_{ji} = \mu_{ij}^*$, ${\bf f}$ is a  symmetric
 matrix,  and generation and color indices are understood.

It can be seen at once from the first two  terms in (\ref{superpot}) 
 that it is impossible to break L-R symmetry with
 such a simple superpotential. The minimum will occur for vanishing vevs 
of $\Delta_c, \bar \Delta_c, \Delta$ and $\bar\Delta $. 
It is clear that the D-term potential vanishes too for the vanishing 
vevs. The addition of soft terms is easily shown to be of no help, since
 the self couplings of the triplet fields have fixed values given 
by the gauge couplings.
Parity cannot be broken in the minimal renormalizable model. 

One can think of two ways out of this problem. The first is to enlarge the
 Higgs sector.
 It was suggested by Kuchimanchi and Mohapatra \cite{km94} to introduce
 a parity-odd
 singlet, coupled appropriately to the triplet fields so as to ensure symmetry 
breaking. However, it was noticed immediately  
that the theory has a set of degenerate
minima connected by a flat direction, all of them breaking parity. 
The problem appears when soft supersymmetry  
breaking terms are switched on: the degeneracy is lifted, but the global
 minimum that results happens to  break electromagnetic charge. Because 
of the flat direction connecting the minima, there is no hope that the 
field remains in the phenomenologically acceptable vacuum, it simply rolls
 down to the global minimum after supersymmetry is softly broken. 
The only way to save the model is to {\em assume a low $SU(2)_R$
breaking  scale}, and 
 the price one has to pay is to break R-parity spontaneously.

In a  recent paper \cite{abs}, two of the authors (C.S.A. and G.S.) with
 K. Benakli, have proved that the minimal extension of 
the Higgs sector 
 consists  on the addition of a couple of triplet fields, $\Omega $
(1,3,1,0) and $\Omega_c$ (1,1,3,0), instead of the singlet. In this model 
the breaking of $SU(2)_R$ is 
achieved in two stages, passing through an intermediate phase
 $SU(2)_L\times U(1)_R\times U(1)_{B-L}$, and breaking $U(1)_R\times U(1)_{B-l}
 $ at a lower scale. This type of low $B-L$ models 
are interesting in their own right, and considered a number of times in 
the literature.
It turns out that this theory
 contains in fact only one parity-breaking minimum, that also
 preserves electromagnetic charge, 
and reduces to the Minimal Supersymmetric Standard Model (MSSM)
 with R-parity.

The second possible way of saving the minimal model is to add 
non-renormalizable terms, while keeping the minimal Higgs content. 
This possibility was suggested in  \cite{mr96b} where non-renormalizable
soft terms  were used to favor the charge-preserving minimum.
However, no systematic study of the effects of the 
non-renormalizable interactions in the superpotential was carried out.
Another example of the use of non-renormalizable terms in B-L models was 
given in \cite{m96}, although not in a manifestly Left-Right symmetric
 model.
We show in the next section how the
addition of non-renormalizable terms suppressed by a high scale 
$M \sim M_{Planck}$, with the field content given by 
(\ref{matter}),(\ref{higgs1}),(\ref{higgs2}) suffices to ensure the correct
 pattern of symmetry breaking. 

\vspace{0.3cm}
{\bf C. The Minimal Non-Renormalizable  Model.} \hspace{0.5cm}
Consider the superpotential (\ref{superpot}). At a first stage, one can 
ignore the terms involving the bidoublet fields $\Phi_i$, that is, we 
can take  a $SU(2)_R$-breaking scale $M_R >> M_W, M_S$. The most general 
 superpotential including non-renormalizable dimension four operators 
that one can write becomes

\begin{eqnarray}
W_{nr} &=&  m ({\rm Tr} \Delta \bar \Delta +
 {\rm Tr} \Delta_c {\bar \Delta}_c)
+i {\bf f} ( L^T \tau_2 \Delta L+ L_c^T\tau_2 \Delta_c L_c )
\nonumber \\
& & + {a \over 2 M}\left[ ({\rm Tr} \Delta \bar\Delta)^2
 + ({\rm Tr} \Delta_c {\bar\Delta}_c)^2 \right]+ {c \over M}{\rm Tr} \Delta \bar\Delta {\rm Tr} \Delta_c \bar\Delta_c \nonumber \\
& & + {b \over 2 M} \left[{\rm Tr} \Delta^2 {\rm Tr} \bar \Delta^2 +
{\rm Tr} \Delta_c^2 {\rm Tr} \bar \Delta_c^2 \right] \nonumber \\
& & + {1 \over M} \left[d_1{\rm Tr} \Delta^2 {\rm Tr}  \Delta_c^2 +
d_2 {\rm Tr} \bar \Delta^2 {\rm Tr} \bar \Delta_c^2 \right] 
\label{nonsuperpot}
\end{eqnarray}
where we assume $M\sim M_{Planck} \simeq 10^{19} GeV$ and
 for simplicity we have taken the couplings to be real.
In the above, we keep the left-handed fields 
since we have to show that parity can be broken spontaneously 
and at the same time we wish to know whether R-parity is broken or not.

The set of minima of the theory are to be determined by imposing the 
vanishing of both $F$ and $D$ terms.
Our first concern is to make sure that these minima are isolated, 
i.e. that there are no flat directions connecting the phenomenologically
allowed minimum with any other  non-physical one.
 Then it would be an easier task to 
prove that the desired 
 minimum exists.
 Above the scale of supersymmetry 
breaking all the minima are  degenerate, therefore we will be concerned
 with potentially dangerous tunneling to physically unacceptable minima
 only at scales below $M_S$. We will finally argue that tunneling at this 
scale is highly suppressed.

The basic result governing the minimization of potentials in globally
supersymmetric 
theories \cite{lt96} is that the space of $D$-flat
vevs may be coordinatized by the set of holomorphic gauge invariants formed 
from the chiral multiplets. The space of flat directions  will be
spanned by the subset of these holomorphic invariants that cannot be 
determined by imposing the $F$-flat conditions. To find this subset in our 
case, we start by considering the $F$ equations for
the left-handed fields  $\Delta$, $\bar\Delta$ and $L$

\begin{eqnarray}
F_{\bar\Delta} &=& (m  + {a \over M} {\rm Tr}\,\Delta\bar\Delta + 
{c\over M}{\rm Tr}\,\Delta_c \bar\Delta_c)\,\Delta \nonumber \\
& &+( {b\over M} {\rm Tr}\,\Delta^2 + {d_1\over M}{\rm Tr}\,\Delta_c^2 )
\, \bar\Delta =0 \nonumber  \\ 
F_\Delta &=&(m  + {a \over M} {\rm Tr}\,\Delta\bar\Delta + 
{c\over M}{\rm Tr}\,\Delta_c \bar\Delta_c)\,\bar \Delta \nonumber \\
& &+( {b\over M} {\rm Tr}\,\bar \Delta^2 + {d_2\over M}{\rm Tr}\,\bar
 \Delta_c^2 )\, \Delta + i F \tau_2 L L^T
=0
\nonumber \\
F_{L} &=& 2 i {\bf f} \tau_2 \Delta L = 0
\label{fflat}
\end{eqnarray}
Here,  we consider for simplicity the case of only one generation
of  leptons. The extension to the realistic multi-generation case is
straightforward.

Clearly, there exists a solution $\langle \Delta\rangle = \langle 
\bar \Delta\rangle = \langle L \rangle = 0$. 
Imposing this condition, we
 are left only with the following holomorphic invariants

\begin{eqnarray}
x_1 &=& {\rm Tr}\,\Delta_c\bar\Delta_c , \quad
x_2 = {\rm Tr}\,\Delta_c^2 {\rm Tr}\,\bar \Delta_c^2 ,  \nonumber \\
x_3 &=& L_c^T \tau_2 \Delta_c L_c , \quad
x_4 = L_c^T \tau_2 \bar \Delta_c L_c {\rm Tr}\,\Delta_c^2 
\label{inv}
\end{eqnarray}

and the $F$-flat conditions
\begin{eqnarray}
F_{\bar\Delta_c} &=& (m  + {a \over M} {\rm Tr}\,\Delta_c\bar\Delta_c)
\,\Delta_c +( {b\over M} {\rm Tr}\,\Delta_c^2 )
\, \bar\Delta_c =0 \nonumber  \\ 
F_{\Delta_c} &=&(m  + {a \over M} {\rm Tr}\,\Delta_c\bar\Delta_c)
\,\bar \Delta_c +( {b\over M} {\rm Tr}\,\bar \Delta_c^2  )\, \Delta_c 
\nonumber \\
& & + i F \tau_2 L_c L_c^T =0
\nonumber \\
F_{L_c} &=& 2 i {\bf f} \tau_2 \Delta_c L_c = 0
\label{fflatc}
\end{eqnarray}

As can be seen immediately, $x_3, x_4$  are made to vanish 
using $F_{L_c}$. It is also straightforward to convince oneself that  
using ${\rm Tr}\, \bar\Delta_c F_{\bar\Delta_c} = 0$ and ${\rm Tr}\, 
\Delta_c F_{\bar\Delta_c} = 0$ the remaining invariants $x_1, x_2$ are 
determined.

It can be shown that equations (\ref{fflatc}) admit in fact two solutions. 
With a definition of electric charge
\begin{equation}
Q = T_{3 L} + T_{3 R} + {1 \over 2} (B-L)
\end{equation}
the  one of interest is 

\begin{equation}
 \langle\Delta^c \rangle  =d  \left ( \begin{array}{cc}
                     \;\; 0 \; \;& \;\; 0 \;\;  \\
                     1 & 0  
\end{array} \right ) ,\quad
\langle \bar\Delta^c \rangle  = d\left ( \begin{array}{cc}
                      \;\;0 \;\; & \;\; 1 \;\;\\
                       0 & 0 \end{array}  \right )  
\label{deltavev}
\end{equation}

with $d = \sqrt{- m M \over a  }$. These vevs break $B-L$ by two
units, and from (\ref{rparity}) we see that  R-parity remains unbroken
at this stage. 

It is an easy task to demonstrate, using $D$ and $F$ terms, that this solution,
 the only one that breaks parity while preserving electromagnetic charge,
 necessarily implies  
\begin{equation}
 \langle L_c \rangle =0
\end{equation}
so that R-parity is preserved in the supersymmetric limit. 
 
Thus we have succeeded in  breaking parity while preserving R-parity.
One can worry that the procedure  above may not be sufficiently general
to ensure that the minimum is indeed isolated, since we have first set
the vevs of the left-handed fields to zero and then required that the
  minimum be isolated in the restricted space of vevs parametrized by
 the right handed gauge invariants. 

To ensure that  flat directions do not run through the parity breaking
minimum, we perturb the vevs  of all fields 
$\Delta,{\bar{\Delta}},\Delta_c,{\bar{\Delta}}_c, L_i, L_i^c$ 
(generically denoted $\psi$)
 by an arbitrary small perturbation 
$ \psi = <\psi> +\epsilon {\hat\psi}$.
We then demand that  the conditions for a supersymmetric 
vacuum $F=D=0$ are satisfied order by order in an expansion in powers of 
$\epsilon =0 $.
  If the resulting equations have non trivial
solutions for the ``flat directions'' ${\hat{\psi}}$,  our minimum
is not isolated . 

For the restricted set of fields kept here it is easy to
show that the parity-breaking  minimum  is indeed isolated,
 i.e $\hat\psi =0$.
For instance the $F=0$ equations for $\Delta$ and $\bar\Delta$
at next to lowest order in $\epsilon$ immediately ensure that
${\hat\Delta}$ and $\hat{\bar\Delta}$ are exactly zero, and continuing
one finds that in fact ${\hat\psi}=0$.
When the bidoublet and quark fields are included the analysis is
more challenging.  Although it is easy to show that even in
their presence the left handed triplets do not participate in
any flat direction through the parity-braking  minimum, there may well be
flat directions through the minimum involving the bidoublet and
quark fields. Details of the analysis in the both cases will be
presented elsewhere.

The scale of 
$SU(2)_R$ breaking  $M_R $ is of order  $\sqrt {m M}$. We leave the 
discussion of the phenomenological implications  for the next session. 
We conclude this one
 with some words on the stability of the vacuum. 
The degenerated minima are separated by barriers of order $M_R$. After soft 
terms become relevant, the degeneracy is lifted up to an order $M_S$. It is
therefore  enough to have $M_R >> M_S$ to get a negligible tunneling 
probability. This is precisely what happens in this model, as we discuss now.

\vspace{0.3cm}
{\bf D. Mass spectrum} \hspace{0.5cm}
We have seen that parity is broken at a scale $M_R$ of order $\sqrt{m M}$. 
Now, it is natural to assume $m$ bigger than the electroweak scale,
for otherwise the soft-breaking terms will effectively mimic its role
 \cite{m96}.
 With 
$m {\buildrel{\sim}\over{>}} 100 GeV $ and $M \sim M_{Planck}$,
 we get a right-handed scale $M_R {\buildrel{\sim}\over{>}} 10^{10} GeV$. 

After symmetry breaking, the Higgs fields get masses through the vev
 of the right-handed triplets in the usual way. However, in some cases
 the mass terms arise form the non-renormalizable terms, thus some 
particles get only a small mass of order $m$. This is the case with the 
left-handed triplets $\Delta$ and $\bar \Delta$, and with the two
 double-charged fields and one of the neutral combinations 
 in $\Delta_c$ and $\bar\Delta_c$.
The remaining fields in $\Delta_c$ and $\bar \Delta_c$
will have  masses of order the $M_R$ scale.

The bidoublet deserves a particular attention. Namely, the
 non-renormalizable superpotential including $\Phi$ will have terms
 of the form

\begin{eqnarray}
W(m_\Phi) &=&  \mu_{ij} {\rm Tr}\,  \tau_2 \Phi^T_i \tau_2 \Phi_j 
+ + {\alpha_{ij} \over M} {\rm Tr}\,  \tau_2 \Phi^T_i 
\bar \Delta_c \Delta_c\tau_2 \Phi_j 
\nonumber \\
 & &+ {\beta_{ij} \over M} {\rm Tr}\,  \tau_2 \Phi^T_i \tau_2 \Phi_j 
 {\rm Tr}\,\Delta_c \bar\Delta_c 
\end{eqnarray}
When $\Delta_c, \bar\Delta_c$ get the vevs (\ref{deltavev}), the
mass terms for $\Phi$ read
\begin{equation}
W(m_\Phi) = \mu_{ij}' {\rm Tr}\,  \tau_2 \Phi^T_i \tau_2 \Phi_j
 + {m \over 2 a} \alpha_{ij} {\rm Tr}\,  \tau_2 \Phi^T_i \tau_3 \tau_2 \Phi_j
\end{equation}
with $\mu_{ij}' = \mu_{ij} + m
(\alpha_{ij}+2 \beta_{ij})/ 2 a $. Thus the two left-handed 
doublets in each bidoublet get split, one of them acquiring a mass of 
order $m$, and the other  (after the usual fine-tuning of the MSSM)
 a mass of the order of the electroweak scale. 

In other words, the minimal L-R model will reduce to the MSSM only below the
  scale $m$.

\vspace{0.3cm}

{\bf E. See-saw mechanism}\hspace{0.5cm}
In the supersymmetric version of Left-Right theories, the see-saw mechanism 
can have novel features. This has been noticed in Ref. \cite{abs}, for the
 model with a low $B-L$ scale cited above. In that case, the $\Delta$ field 
coupling to the left-handed neutrino does not acquire a vev, in sharp 
contrast with the non-supersymmetric case \cite{ms81}. The see-saw mechanism is then
 said to be ``clean'', in the sense that it takes its  canonical form. 

This is not the case however in the non-renormalizable minimal model. Namely,
 the bidoublet superpotential will have terms like
\begin{equation}
W_{NR}(\Phi) = ...+{\eta_{ij} \over M}{\rm Tr}\, 
\tau_2 \Phi^T_i \Delta_c \tau_2 \Phi_j \Delta  +... 
\end{equation}
which will give rise to terms linear in $\Delta$ after parity breaking, 
of the order $\sqrt{m/M}$. Such tadpole term  will force $\Delta $ to
 get a vev after electroweak breaking 
$\langle \Delta\rangle \sim M_W^2/M_R$, which is precisely the situation
 one encounters on the non-supersymmetric version of the theory. This has
 an  impact on neutrino masses, and  provides
 an important distinction from the renormalizable version of supersymmetric
 Left-Right models \cite{abs}.

\vspace{0.3cm}

{\bf  R-parity conservation}\hspace{0.5cm} As we have seen, at the large 
scale, charge conservation demands also conservation of R-parity.
The question is what happens after the heavy fields are integrated out 
and the soft supersymmetry  breaking terms are switched on. 
Here the analysis proceeds 
completely along the lines of Ref. \cite{abs}. Since $M_R$ is very large,
the breakdown of R-parity at low energies would imply an almost-massless 
majoron coupled to the Z-boson, which is ruled out experimentally. This is 
one of the central aspects of supersymmetric left-right theories with large
 $M_R$: R-parity is an exact symmetry of the low-energy effective theory.
 This has well-known important phenomenological and cosmological
  implications. In particular, the lightest supersymmetric partner must 
be stable, becoming a natural dark matter candidate.

\vspace{0.3cm}

{\bf Summary and Outlook}\hspace{0.5cm} Left-Right symmetry (or $B-L$) 
provides a natural gauge principle rationale for R-parity, and thus offers a
framework for the study of the predictivity of its breaking. It also plays
 an important role in understanding the smallness of strong CP violation. On 
the other hand, it turned out surprisingly hard to construct a realistic 
supersymmetric Left-Right model and it was claimed that the minimal such 
theory cannot work (unless $M_R \leq M_S$).

However, we find out that the simple inclusion of non-renormalizable $d=4$
 terms in the superpotential, even if cut-off by $M_{Planck}$, leads to 
a perfectly consistent model with the spontaneous breakdown of parity.

Our predictions are

1) a number of charged and doubly-charged Higgs scalars with a mass 
$m\simeq M_R^2/M_{Planck}$. Thus, even for a large $M_R$ in the
 intermediate scale $10^{10}-10^{12} $ GeV, interesting for neutrino physics, 
these new particles can be found in the near future experiments.
 This is the crucial prediction.

2) R-parity  remains an exact symmetry of the low-energy theory.

3) the see-saw mechanism takes a similar form  as in the 
non-supersymmetric models, and this  is in sharp contrast with the 
renormalizable version.

We leave the last word to experiment.

\vspace{0.5cm}
We wish to thank Andrija Ra\v{s}in for his interest in this project,
for many valuable discussions and a careful reading of the manuscript.
We are also grateful to Karim Benakli for discussions in the early 
stages of this 
work and for his interest and useful comments. A.M wishes to thank L. Prada
for primordial support.

The work of G.S. is supported in part  by EEC 
under the TMR contract ERBFMRX-CT960090.

\end{document}